\documentclass[aps,prl,twocolumn,superscriptaddress,amsmath]{revtex4-1}
\usepackage{graphicx,color}

\begin{document}

\title{Vulnerability and co-susceptibility determine the size of network cascades}

\author{Yang Yang}
\affiliation{Department of Physics and Astronomy, Northwestern University, Evanston, IL 60208, USA}

\author{Takashi Nishikawa}
\affiliation{Department of Physics and Astronomy, Northwestern University, Evanston, IL 60208, USA}
\affiliation{Northwestern Institute on Complex Systems, Northwestern University, Evanston, IL 60208, USA}

\author{Adilson E. Motter}
\affiliation{Department of Physics and Astronomy, Northwestern University, Evanston, IL 60208, USA}
\affiliation{Northwestern Institute on Complex Systems, Northwestern University, Evanston, IL 60208, USA}

\begin{abstract}
In a network, a local disturbance can propagate and eventually cause a substantial part of the system to fail, in cascade events that are easy to conceptualize but extraordinarily difficult to predict. Here, we develop a statistical framework that can predict cascade size distributions by incorporating two ingredients only: the vulnerability of individual components and the co-susceptibility of groups of components (i.e., their tendency to fail together). Using cascades in power grids as a representative example, we show that correlations between component failures define structured and often surprisingly large {\it groups} of co-susceptible components. Aside from their implications for blackout studies, these results provide insights and a new modeling framework for understanding cascades in financial systems, food webs, and complex networks in general.
\end{abstract}

\maketitle 

The stability of complex networks is largely determined by their ability to operate close to equilibrium---a condition that can be compromised by relatively small perturbations that can lead to large cascades of failures.
Cascades are responsible for a range of network phenomena, from power blackouts \cite{hines2009large} and air traffic delay propagation~\cite{fleurquin2013} to secondary species extinctions \cite{sahasrabudhe2011rescuing,eco:11} and large social riots~\cite{watts2002,brummitt2015}.
Evident in numerous previous modeling efforts~\cite{bak_1987, kinney_2005, buldyrev_2010, watts2002, goh_2003, motter_2004, dobson2007complex, brummitt2015, sahasrabudhe2011rescuing, eco:11} is that dependence between components is the building block of these self-amplifying processes and can lead to correlations among eventual failures in a cascade.

A central metric characterizing a cascade is its size.
While the suitability of a size measure depends on the context and purpose, a convenient measure is the number of network components (nodes or links) participating in the cascade (e.g., failed power lines, delayed airplanes, extinct species).
Since there are many known and unknown factors that can affect the details of cascade dynamics, the main focus in the literature has been on characterizing the statistics of cascade sizes rather than the size of individual events.
This leads to a fundamental question: what determines the distribution of cascade sizes?

\begin{figure}[t] 
\hspace{-2.8mm}\includegraphics[width=3.5in]{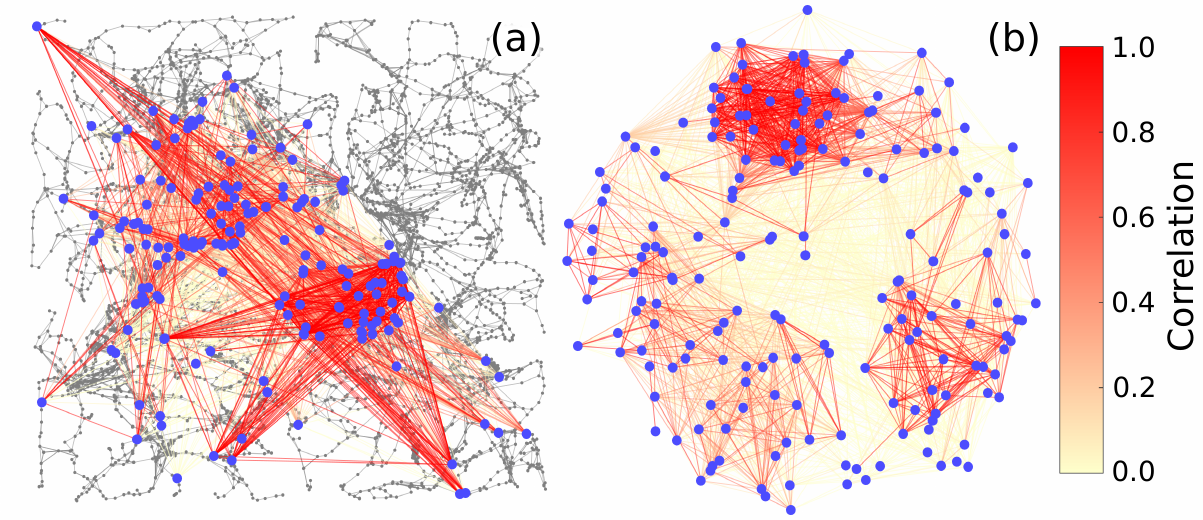}\\[-2mm]
\caption{
Example of co-susceptibility of cascading failures in a power grid.
(a) Color-coded network of positive correlations between failures of transmission lines (blue dots) in the Texas network.
In the gray background network, the nodes represent (all) transmission lines, while each link represents a direct physical connection through a common substation.
(b)~Network in (a) after a correlation-based repositioning of the nodes.}
\label{fig1}
\end{figure}

In this Letter, we show that cascading failures (and hence their size distributions) are often determined primarily by two key properties associated with failures of the system components: the {\it vulnerability}, or the failure probability of each component, and the 
{\it co-susceptibility}, or the tendency of a group of components to fail together.
The latter is intimately related to pairwise correlations between failures, as we will see below.
We provide a concrete algorithm for identifying groups of co-susceptible components for any given network.
We demonstrate this using the representative example of 
cascades of overload failures in power grids 
(Fig.~\ref{fig1}).
Based on our findings, we develop the {\it co-susceptibility model}---a statistical modeling framework capable of accurately predicting the distribution of cascade sizes, depending solely on the vulnerability and co-susceptibility of component failures.

We consider a system of $n$ components subject to cascading failures, in which a set of initial component failures can induce a sequence of failures in other components.
Here we assume that the initial failures and the propagation of failures can be modeled as stochastic and deterministic processes, respectively (although the framework also applies if the propagation or both are stochastic).
Thus, the cascade size $N$, defined here as the total number of components that fail after the initial failures, is a random variable that can be expressed as 
\begin{equation}\label{eqn2}
N = \sum_{\ell=1}^n F_\ell, 
\end{equation}
where $F_\ell$ is a binary random variable representing the failure status of component $\ell$ (i.e., $F_\ell = 1$ if component $\ell$ fails during the cascade, and $F_\ell = 0$ otherwise). 
While the $n$ components may be connected by physical links, a component may fail as the cascade propagates even if none of its immediate neighbors have failed~\cite{brummitt2015,Witthaut:2015,dobson_2016}.
For example, in the case of cascading failures of transmission lines in a power grid, the failure of one line can cause a reconfiguration of power flows across the network that leads to the overloading and subsequent failure of other lines away from the previous failures~\cite{dobson_2016,anghel2007stochastic}.

A concrete example network we analyze throughout this Letter using the general setup above is the Texas power grid, for which we have 24 snapshots, representing on- and off-peak power demand in each season of three consecutive years.
Each snapshot comprises the topology of the transmission grid, the capacity threshold of each line, the power demand of each load node, and the power supply of each generator node (extracted from the data reported to FERC~\cite{FERC}).
For each snapshot we use a physical cascade model to generate $K=5{,}000$ cascade events.
In this model (which is a variant of that in Ref.~\cite{anghel2007stochastic} with the power re-balancing scheme from Ref.~\cite{Hines2011}),
an initial perturbation to the system (under a given condition) is modeled by the removal of a set of randomly selected lines. 
A cascade following the initial failures is then modeled as an iterative process.
In each step, power flow is redistributed according to Kirchhoff's law and might therefore cause some lines to be overloaded and removed (i.e., to fail) due to overheating.
The temperature of the transmission lines is described by a continuous evolution model and the overheating threshold for line removal is determined by the capacity of the line~\cite{anghel2007stochastic}.
When a failure causes part of the grid to be disconnected, we re-balance power supply and demand under the constraints of limited generator capacity~\cite{Hines2011}.
A cascade stops when no more overloading occurs, and we define the size $N$ of the cascade as the total number of removed lines (excluding the initial failures).
This model~\cite{source-code}, accounting for several physical properties of failure propagation, sits relatively high in the hierarchy of existing power-grid cascade models~\cite{opaModel,henneaux2016,dobson2012_vulnerability,hines_2015}, which ranges from the most detailed engineering models to simplest graphical or stochastic models.
The model has also been validated against historical data~\cite{Yang:2016}.

In general, mutual dependence among the variables $F_\ell$ may be necessary to explain the distribution of the cascade size $N$.
We define the {\it vulnerability} $p_\ell \equiv \langle F_\ell \rangle$ of component $\ell$ to be the probability that this component fails in a cascade event (including events with $N=0$).
If the random variables $F_\ell$ are uncorrelated (and thus have zero covariance), then $N$ would follow   Poisson's binomial distribution~\cite{Wang:1993}, with average $\tilde{\mu}=\sum_\ell p_\ell$ and variance $\tilde{\sigma}^2=\sum_\ell p_\ell(1-p_\ell)$.
However, the actual variance $\sigma^2$ of $N$ observed in the cascade-event data is significantly larger than the corresponding value $\tilde{\sigma}^2$ under the no-correlation assumption for all $24$ snapshots of the Texas power grid (with the relative difference, $\bar{\sigma}^2\equiv(\sigma^2 - \tilde{\sigma}^2)/\tilde{\sigma}^2$, ranging from around $0.18$ to nearly $39$).
Thus, the mutual dependence must contribute to determining the distribution of $N$ in these examples.

Part of this dependence is captured by the correlation matrix $C$, whose elements are the pairwise Pearson correlation coefficients among the failure status variables $F_\ell$.
When the correlation matrix is estimated from cascade-event data, it has noise due to finite sample size, which we filter out using the following procedure.
First, we standardize $F_\ell$ by subtracting the average and dividing it by the standard deviation.
According to random matrix theory, the probability density of eigenvalues of the correlation matrix computed from $K$ samples of $T$ independent random variables follow the Marchenko-Pastur distribution~\cite{mehta2004random}, $\rho(\lambda) = K\sqrt{(\lambda_{+} - \lambda)(\lambda - \lambda_{-})}/(2 \pi \lambda T)$, where $\lambda_{\pm}  = \bigl[1 \pm \sqrt{T/K}\,\bigr]^2$. 
Since those eigenvalues falling between $\lambda_{-}$ and $\lambda_{+}$ can be considered contributions from the noise, the sample correlation matrix $\widehat{C}$ can be decomposed as $\widehat{C}  = \widehat{C}^{(\text{ran})}+\widehat{C}^{(\text{sig})}$, where $\widehat{C}^{(\text{ran})}$ and $\widehat{C}^{(\text{sig})}$ are its random and significant parts, respectively, which can be determined from the eigenvalues and the associated eigenvectors~\cite{MacMahon:2015}.
In the network visualization of Fig.~\ref{fig1}(a), we show the correlation coefficients $\widehat{C}_{\ell\ell'}^{\text{(sig)}}$ between components $\ell$ and $\ell'$ estimated from the cascade-event data for the Texas grid under the 2011 summer on-peak condition.
Note that we compute correlation only between those components that fail more than once in the cascade events.
As this example illustrates, we observe no apparent structure in a typical network visualization of these correlations.
However, as shown in Fig.~\ref{fig1}(b), after repositioning the nodes based on correlation strength, we can identify clusters of positively and strongly correlated components---those that tend to fail together in a cascade.

To more precisely capture this tendency of simultaneous failures, we define a notion of {\it co-susceptibility}: a given subset of $m$ components $\mathcal{I}\equiv\{\ell_1,\ldots,\ell_m\}$ is said to be co-susceptible if 
\begin{equation}\label{eqn:co-sus}
\gamma_\mathcal{I} \equiv \frac{\langle N_\mathcal{I} \vert N_\mathcal{I}\neq0 \rangle - \bar{n}_\mathcal{I}}{m - \bar{n}_\mathcal{I}} > \gamma_\text{th},
\end{equation}
where $N_\mathcal{I}\equiv\sum_{j=1}^m F_{\ell_j}$ is the number of failures in a cascade event among the $m$ components, $\langle N_\mathcal{I} \vert N_\mathcal{I}\neq0 \rangle$ denotes the average number of failures among these components given that at least one of them fails, $\bar{n}_\mathcal{I}\equiv \sum_{j=1}^m p_{\ell_j}/\bigl[1-\prod_{k=1}^m(1-p_{\ell_k})\bigr] \ge 1$ is the value $\langle N_\mathcal{I} \vert N_\mathcal{I}\neq0 \rangle$ would take if $F_{\ell_1},\ldots,F_{\ell_m}$ were independent.
Here we set the threshold in Eq.~\eqref{eqn:co-sus} to be $\gamma_\text{th}=\sigma_{N_\mathcal{I}}/(m - \bar{n}_\mathcal{I})$, where $\sigma_{N_\mathcal{I}}^2 \equiv \sum_{j=1}^m p_{\ell_j}(1-p_{\ell_j})/\bigl[1-\prod_{k=1}^m(1-p_{\ell_k})\bigr] - \bar{n}_\mathcal{I}^2 \prod_{k=1}^m(1-p_{\ell_k})$ is the variance of $N_\mathcal{I}$ given $N_\mathcal{I}\neq0$ for statistically independent $F_{\ell_1},\ldots,F_{\ell_m}$.
By definition, the co-susceptibility measure $\gamma_\mathcal{I}$ equals zero if $F_{\ell_1},\ldots,F_{\ell_m}$ are independent.
It satisfies $-(\bar{n}_\mathcal{I}-1)/(m-\bar{n}_\mathcal{I}) \le \gamma_\mathcal{I} \le 1$, where the (negative) lower bound is achieved if multiple failures never occur and the upper bound is achieved if all $m$ components fail whenever one of them fails.
Thus, a set of co-susceptible components are characterized by significantly larger number of simultaneous failures among these components, relative to the expected number for statistically independent failures.
While $\gamma_\mathcal{I}$ can be computed for a given set of components, identifying sets of co-susceptible components in a given network from Eq.~\eqref{eqn:co-sus} becomes infeasible quickly as $n$ increases due to combinatorial explosion.

\begin{figure}[t!] 
\includegraphics[width=8.5cm]{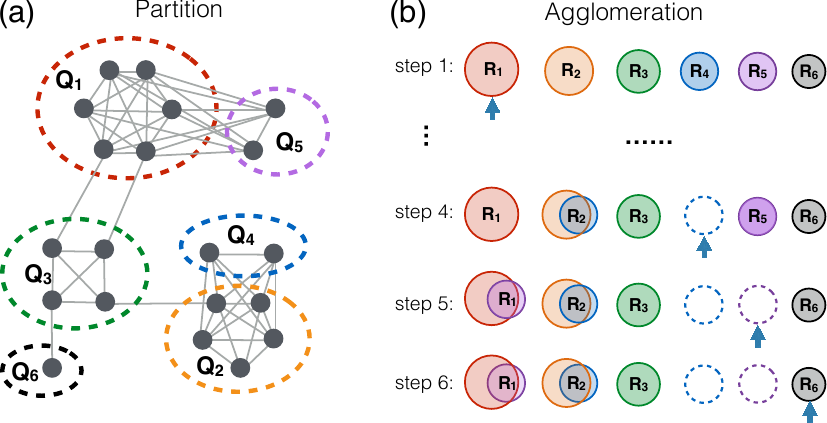}\\[-2mm]
\caption{
Two-stage algorithm for identifying sets of co-susceptible components.
(a) First stage, in which we partition the reduced unweighted correlation graph into multiple cliques
$\{Q_k\}$, and index the cliques in descending order of size.
(b) Second stage, in which we find the co-susceptible groups $\{R_k\}$ by recursively agglomerating cliques that have enough connections.
\label{fig2}}
\end{figure}

Here we propose an efficient two-stage algorithm for identifying co-susceptible components~\cite{source-code}.
The algorithm is based on partitioning and agglomerating the vertices of the auxiliary graph $G_0$ in which vertices represent the components that fail more than once in the cascade-event data, and (unweighted) edges represent the dichotomized correlation between these components.
Here we use $\widehat{C}^{(\text{sig})}_{\ell \ell'}>0.4$ as the criteria for having an edge between vertices $\ell$ and $\ell'$ in $G_0$.
In the first stage [illustrated in Fig.~\ref{fig2}(a)], $G_0$ is divided into non-overlapping cliques---subgraphs within which any two vertices are directly connected---using the following iterative process.
In each step $k=1,2,\ldots$, we identify a clique of the largest possible size (i.e., the number of vertices it contains), denote this clique as $Q_k$, remove $Q_k$ from the graph $G_{k-1}$, and then denote the remaining graph by $G_k$.
Repeating this step for each $k$ until $G_k$ is empty, we obtain a sequence $Q_1, Q_2, \ldots, Q_m$ of non-overlapping cliques in $G_0$, indexed in the order of non-increasing size. 
In the second stage, we agglomerate these cliques, as illustrated in Fig.~\ref{fig2}(b).
Initially, we set $R_k = Q_k$ for each $k$. 
Then, for each $k=2,3,\ldots,m$, we either move all the vertices in $R_k$ to the largest group among  $R_1,\cdots, R_{k-1}$ for which at least $80\%$ of all the possible edges between that group and $R_k$ actually exist, or we keep $R_k$ unchanged if no group satisfies this criterion.
Among the resulting groups, we denote those groups whose size is at least three by $R_1,R_2,\ldots,R_{m'}$, $m'\le m$.
A key advantage of our method over applying community-detection algorithms~\cite{MacMahon:2015} is that the edge density threshold above can be optimized for the accuracy of cascade size prediction.

\begin{figure}[t!]
\begin{center}
\includegraphics[width=\columnwidth]{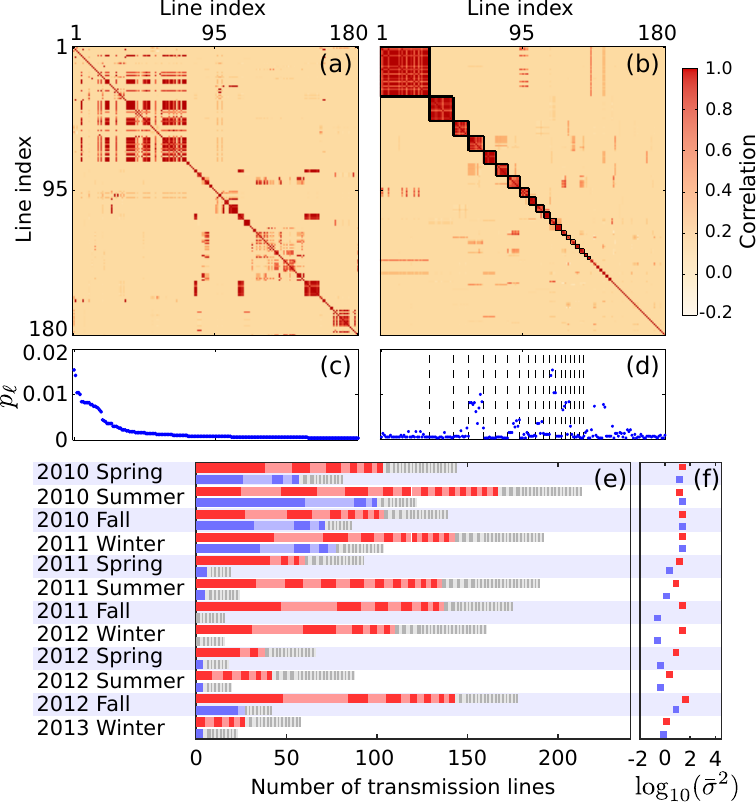}
\end{center}
\vspace{-5mm}
\caption{
Correlations between transmission line failures in the Texas power grid.
(a--d) Estimated correlation matrix $\widehat{C}^{\text{(sig)}}$ [(a),(b)] and vulnerabilities $p_\ell$ [(c),(d)] under the 2011 summer on-peak condition. 
In (a) and (c), the transmission lines are indexed so that $p_\ell$ is decreasing in $\ell$.
In (b) and (d), they are indexed so that the sets of co-susceptible lines appear as diagonal blocks.
(e) Sizes of the groups of co-susceptible transmission lines, indicated by the lengths of the individual segments of the red portion of the top bar (on-peak snapshot) and of the blue portion of the bottom bar (off-peak snapshot) for each season.
The gray portion of each bar corresponds to groups of fewer than three lines among groups $R_k$.
(f) Relative difference $\bar{\sigma}^2$ between the variance of the cascade size and its counterpart under the no-correlation assumption.
\label{fig:correlation}}
\end{figure}

We test the effectiveness of our general algorithm on the Texas power grid.
As Figs.~\ref{fig:correlation}(a) and \ref{fig:correlation}(b) show, the block-diagonal structure of $\widehat{C}^{\text{(sig)}}$ indicating high correlation within each group and low correlation between different groups becomes evident when the components are reindexed according to the identified groups.
We note, however, that individual component vulnerabilities do not necessarily correlate with the co-susceptibility group structure [see Fig.~\ref{fig:correlation}(d), in comparison with Fig.~\ref{fig:correlation}(c)].
We find that the sizes of the groups of co-susceptible components vary significantly across the $24$ snapshots of the Texas power grid, as shown in Fig.~\ref{fig:correlation}(e).
The degree of co-susceptibility, as measured by the total number of co-susceptible components, is generally lower under an off-peak condition than the on-peak counterpart [Fig.~\ref{fig:correlation}(e)].
This is consistent with the smaller deviation from the no-correlation assumption observed in Fig.~\ref{fig:correlation}(f), where this deviation is measured by the relative difference in the variance, $\bar{\sigma}^2$ (defined above).
Since high correlation within a group of components implies a high probability that many of them fail simultaneously, the groups identified by our algorithm tend to have high values of $\gamma_\mathcal{I}$.
Indeed, our calculation shows that Eq.~\eqref{eqn:co-sus} is satisfied for all the $171$ co-susceptible groups found in the $24$ snapshots.

\begin{figure}[t!] 
\includegraphics[width=\columnwidth]{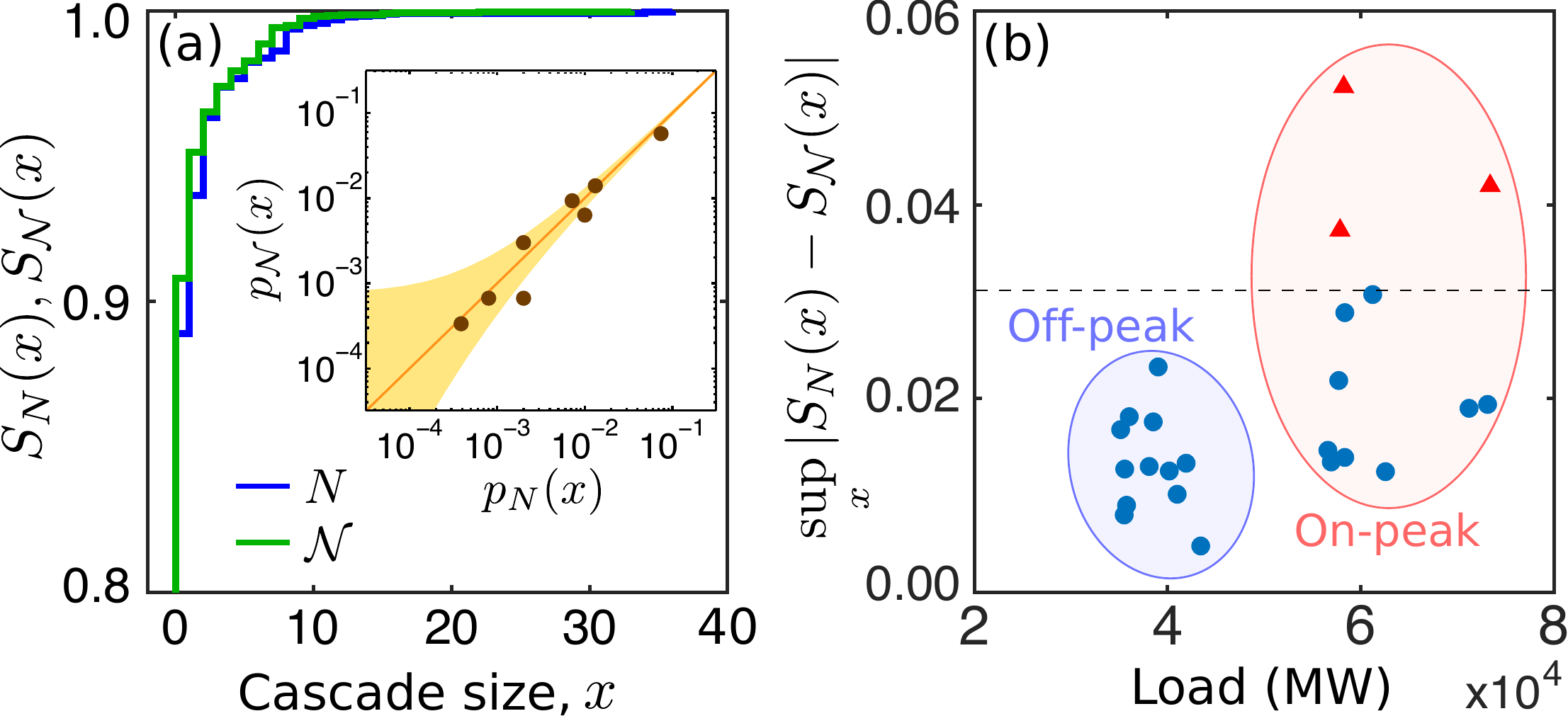}
\vspace{-0.5cm}
\caption{
Validation of the co-susceptibility model.
(a) Cumulative distributions $S_N(x)$ and $S_{\cal N}(x)$ of cascade sizes $N$ and ${\cal N}$ from the cascade-event data and from the co-susceptibility model, respectively, under the 2011 summer on-peak condition.
Inset: Binned probabilities from the two distributions plotted against each other.
The shaded area indicates the 95\% confidence interval.
(b) Distance between two distributions as a function of the total load in megawatts (MW) for all $24$ snapshots.
}\label{fig:model-validation}
\end{figure}

Given the groups of components generated through our algorithm, 
the {\it co-susceptibility model} is defined as the set of binary random variables ${\cal F}_\ell$ (different from $F_\ell$) following the dichotomized correlated Gaussian distribution~\cite{emrich1991method,macke2009generating} whose marginal probabilities (i.e., the probabilities that ${\cal F}_\ell=1$) equal the estimates of $p_\ell$ from the cascade-event data and whose correlation matrix ${\cal C}$ is given by
\begin{equation}
{\cal C}_{\ell\ell'} = \begin{cases}
\widehat{C}^{(\text{sig})}_{\ell\ell'} & \text{if $\ell,\ell'\in R_k$ for some $k\le m'$,}\\
0 & \text{otherwise.}
\end{cases}
\end{equation}
We are thus approximating the correlation matrix $C$ by the block diagonal matrix ${\cal C}$, where the blocks correspond to the sets of co-susceptible components.
In terms of the correlation network, this corresponds to using only those links within the same group of co-susceptible components for predicting the distribution of cascade sizes.
Since individual groups are assumed to be uncorrelated, this can be interpreted as model dimensionality reduction, in which the dimension reduces from $n$ to the size of the largest group.
We sample ${\cal F}_\ell$ using the code provided in Ref.~\cite{macke2009generating}.
In this implementation, the computational time for sampling scales with the number of variables with an exponent of $3$, so factors of $2.0$ to $15.2$ in dimensionality reduction observed for the Texas power grid correspond to a reduction of computational time by factors of more than $8$ to more than $3{,}500$.

We now validate the co-susceptibility model for the Texas grid.
We estimate the cumulative distribution function $S_{\cal N}(x)$ of cascade size, ${\cal N}\equiv \sum_\ell {\cal F}_\ell$, using $3{,}000$ samples generated from the model. 
As shown in Fig.~\ref{fig:model-validation}(a), this function matches well with the cumulative distribution function $S_N(x)$ of cascades size $N$ computed directly from the cascade-event data.
This is validated more quantitatively in the inset; the (binned) probability $p_{\cal N}(x)$ that $x \le {\cal N} \le x+\Delta x$ for the co-susceptibility model is plotted against the corresponding probability $p_N(x)$ for the cascade-event data, using a bin size of $\Delta x = N_{\max}/20$, where $N_{\max}$ denotes the maximum cascade size observed in the cascade-event data.
The majority of the points lie within the 95\% confidence interval for $p_{\cal N}(x)$, computed using the estimated $p_N(x)$.
To validate the co-susceptibility model across all 24 snapshots, we use the Kolmogorov-Smirnov (KS) test~\cite{massey1951kolmogorov}.
Specifically, for each snapshot we test the hypothesis that the samples of ${\cal N}$ and the corresponding samples of $N$ are from the same distribution. 
Figure~\ref{fig:model-validation}(b) shows the measure of distance between two distributions, $\sup_x |S_N(x)-S_{\cal N}(x)|$, which underlies the KS test, as a function of the total amount of electrical load in the system.
We find that the null hypothesis cannot be rejected 
at the 5\% significance level
for most of the cases we consider [$21/24=87.5$\%, blue dots in Fig.~\ref{fig:model-validation}(b)]; it can be rejected in only three cases (red triangles, above the threshold distance indicated by the dashed line), all corresponding to high stress (i.e., high load) conditions.
We also see that more stressed systems are associated with larger distances between the distributions, and a higher likelihood of being able to reject the null hypothesis.
We believe this is mainly due to higher-order correlations not captured by $p_\ell$ and $C$.

The identification of co-susceptibility as a key ingredient in determining cascade sizes leads to two new questions:
(1) What gives rise to co-susceptibility? 
(2) How to identify the co-susceptible groups?
While the first question opens an avenue for future research, the second question is addressed by the algorithm developed here (for which we provide a ready-to-use software~\cite{source-code}). 
The co-susceptibility model is general and can be used for cascades of any type (of failures, information, or any other spreadable attribute) for which information is available on the correlation matrix and the individual ``failure'' probabilities.
Such information can be empirical, as in the financial data studied in Ref.~\cite{Plerou:2002}, or generated from first-principle models, as in the power-grid example used here.
Our approach accounts for correlations (a strength shared by some other approaches, such as the one based on branching processes~\cite{dobson2012}), and does so from the fresh, network-based perspective of co-susceptibility.
Finally, since co-susceptibility is often a nonlocal effect, our results suggest that we may need nonlocal strategies for reducing the risk of cascading failures, which bears implications for future research.

\begin{acknowledgments}
This work was supported by ARPA-E Award \begin{NoHyper}No.~DE-AR0000702\end{NoHyper}.
\end{acknowledgments}



\begin{thebibliography}{}

\bibitem{hines2009large} 
P.~Hines, J.~Apt, and S.~Talukdar,
Large blackouts in North America: Historical trends and policy implications,
Energ. Policy \textbf{37}, 5249 (2009). 

\bibitem{fleurquin2013}
P.~Fleurquin, J.~J.~Ramasco, and V.~M.~Eguiluz,
Systemic delay propagation in the US airport network,
Sci. Rep. \textbf{3}, 1159 (2013).

\bibitem{eco:11}
J.~A.~Estes, J.~Terborgh, J.~S.~Brashares, M.~E.~Power, J.~Berger, W.~J.~Bond, and D.~A.~Wardle,
Trophic downgrading of planet Earth,
Science \textbf{333}, 301 (2011). 

\bibitem{sahasrabudhe2011rescuing}
S.~Sahasrabudhe and A.~E.~Motter,
Rescuing ecosystems from extinction cascades through compensatory perturbations,
Nature Commun. \textbf{2}, 170 (2011).

\bibitem{watts2002}
D.~J.~Watts,
A simple model of global cascades on random networks,
Proc. Natl. Acad. Sci. USA \textbf{99}, 5766 (2002). 

\bibitem{brummitt2015}
C.~D.~Brummitt, G.~Barnett, and R.~M.~D'Souza,
Coupled catastrophes: sudden shifts cascade and hop among interdependent systems,
J. Roy. Soc. Interface \textbf{12}, 20150712 (2015).

\bibitem{kinney_2005}
R.~Kinney, P.~Crucitti, R.~Albert, and V.~Latora,  
Modeling cascading failures in the North American power grid,
Euro. Phys. J. B \textbf{46}, 101 (2005). 

\bibitem{buldyrev_2010}
S.~V.~Buldyrev, R.~Parshani, G.~Paul, H.~E.~Stanley, and S.~Havlin,
Catastrophic cascade of failures in interdependent networks,
Nature \textbf{ 464}, 1025 (2010). 

\bibitem{goh_2003}
K.~I.~Goh, D.~S.~Lee, B.~Kahng, and D.~Kim, 
Sandpile on scale-free networks,
Phys. Rev. Lett. \textbf{91}, 148701 (2003).

\bibitem{motter_2004}
A.~E.~Motter,  
Cascade control and defense in complex networks,
Phys. Rev. Lett. \textbf{93}, 098701 (2004).

\bibitem{dobson2007complex}
I.~Dobson, B.~A.~Carreras, V.~E.~Lynch, and D.~E.~Newman, 
Complex systems analysis of series of blackouts: Cascading failure, critical points, and self-organization,
Chaos \textbf{17}, 026103 (2007).

\bibitem{bak_1987}
P.~Bak, C.~Tang, and K.~Wiesenfeld,
Self-organized criticality: An explanation of the $1/f$ noise,
Phys. Rev. Lett. \textbf{59}, 381 (1987).

\bibitem{Witthaut:2015}
D.~Witthaut and M.~Timme,
Nonlocal effects and countermeasures in cascading failures,
Phys. Rev. E \textbf{92}, 032809 (2015).

\bibitem{dobson_2016}
I.~Dobson, B.~A.~Carreras, D.~E.~Newman, and J.~M.~Reynolds-Barredo,  
Obtaining statistics of cascading line outages spreading in an electric transmission network from standard utility data,
IEEE T. Power Syst. \textbf{99}, 1 (2016). 

\bibitem{anghel2007stochastic}
M.~Anghel, K.~A.~Werley, and A.~E.~Motter,
Stochastic model for power grid dynamics,
{\em Proc. 40th Int. Conf. Syst. Sci.} HICSS'07, 
Big Island, HI, USA,
Vol. 1, 113 (2007).

\bibitem{FERC}
The data for the snapshots are obtained from Federal Energy Regulatory Commission (FERC) Form 715.

\bibitem{Hines2011}
P.~Hines, E.~Cotilla-Sanchez, and S.~Blumsack,
Topological models and critical slowing down: Two approaches to power system blackout risk analysis,
{\em Proc. 44th Int. Conf. Syst. Sci.} HICSS'11, 
Kauai, HI, USA, 1 (2011).

\bibitem{source-code}
Source code is available for download at \url{https://github.com/yangyangangela/determine_cascade_sizes}

\bibitem{opaModel}
B.~A.~Carreras, D.~E.~Newman, I.~Dobson, and N.~S.~Degala,
Validating OPA with WECC data,
{\em Proc. 46th Int. Conf. Syst. Sci.} HICSS'13, 
Maui, HI, USA, 2197 (2013).

\bibitem{henneaux2016}
P.~Henneaux, P.~E.~Labeau, J.~C.~Maun, and L.~Haarla,
A two-level probabilistic risk assessment of cascading outages,
IEEE T. Power Syst. \textbf{31}, 2393 (2016).

\bibitem{hines_2015}
P.~D.~Hines, I.~Dobson, and P.~Rezaei,  
Cascading power outages propagate locally in an influence graph that is not the actual grid topology,
arXiv:1508.01775 (2016). 

\bibitem{dobson2012_vulnerability}
B.~A.~Carreras, D.~E.~Newman, and I.~Dobson,
Determining the vulnerabilities of the power transmission system,
{\em Proc. 45th Int. Conf. Syst. Sci.} HICSS'12,
Maui, HI, USA, 2044 (2012). 

\bibitem{Yang:2016}
Y.~Yang, T.~Nishikawa, and A.~E.~Motter (to be published).

\bibitem{Wang:1993}
Y.~H.~Wang, 
On the number of successes in independent trials,
Stat. Sinica. {\bf 3}, 295 (1993). 

\bibitem{mehta2004random}
M.~L.~Mehta,
{\em Random Matrices}, 3rd edn.
(Academic Press, 2004).

\bibitem{MacMahon:2015}
M.~MacMahon and D.~Garlaschelli,
Community Detection for Correlation Matrices, 
Phys. Rev. X {\bf 5}, 021006 (2015).

\bibitem{emrich1991method}
L.~J.~Emrich and M.~R.~Piedmonte,
A method for generating high-dimensional multivariate binary variates,
Amer. Statist. \textbf{45}, 302 (1991). 

\bibitem{macke2009generating}
J.~H.~Macke, P.~Berens, A. S.~Ecker, A.~S.~Tolias, and M.~Bethge,
Generating spike trains with specified correlation coefficients,
Neural Comp. \textbf{21}, 397 (2009). 

\bibitem{massey1951kolmogorov}
F.~J.~Massey, Jr.,
The Kolmogorov-Smirnov test for goodness of fit,
J. Amer. Statist. Assoc. \textbf{46}, 68 (1951). 

\bibitem{Plerou:2002}
V.~Plerou, P.~Gopikrishnan, B.~Rosenow, L.~A.~N.~Amaral, T.~Guhr, and H.~E.~Stanley,
Random matrix approach to cross correlations in financial data,
Phys. Rev. E {\bf 65}, 066126 (2002).

\bibitem{dobson2012}
I.~Dobson, 
Estimating the propagation and extent of cascading line outages from utility data with a branching process,
IEEE T. Power Syst. \textbf{27}, 2146 (2012). 

\end{thebibliography}
\end{document}